# ANALYSIS OF DISCRETE SIGNALS WITH STOCHASTIC COMPONENTS USING FLICKER NOISE SPECTROSCOPY

SERGE F. TIMASHEV

*Karpov Institute of Physical Chemistry, Moscow 103064, Russia*

*timashev@cc.nifhi.ac.ru*

YURIY S. POLYAKOV

*USPolyResearch, Ashland, PA 17921, USA*

*ypolyakov@uspolyresearch.com*

**Abstract**

The problem of information extraction from discrete stochastic[1] time series, produced with some finite sampling frequency, using flicker-noise spectroscopy, a general framework for information extraction based on the analysis of the correlation links between signal irregularities and formulated for continuous signals, is discussed. It is shown that the mathematical notions of Dirac $\delta$- and Heaviside $\theta$- functions used in the analysis of continuous signals may be interpreted as high-frequency and low-frequency stochastic components, respectively, in the case of discrete series. The analysis of electroencephalogram measurements for a teenager with schizophrenic symptoms at two different sampling frequencies demonstrates that the "power spectrum" and difference moment contain different information in the case of discrete signals, which was formally proven for continuous signals. The sampling interval itself is suggested as an additional parameter that should be included in general parameterization procedures for real signals.

---

[1] The term "stochastic" in this paper refers to the presence of random variability in the signals of complex systems

## 1. Introduction

Stochastic time and space series of dynamic variables that arise in studies of various natural processes and structure are often an important source of information about the system state and features of its evolution and structure [Réfrégier, 2004]. Therefore, a reliable tool to extract and analyze the information contained in the series could help understand many processes occurring in nature, for example, the preparation stages for a major earthquake or development of a human disease.

The raw data about the dynamics of complex real systems are usually obtained as discrete time series $V(t_k)$ produced with some finite sampling frequency $f_d$, where $t_k$ are time moments separated by a fixed interval $\Delta t = f_d^{-1}$. When extracting information from these discrete time series, one first needs to answer the fundamental question: how complete and reliable is the information contained in the signals recorded with some finite sampling frequency considering that real systems also generate signals at much higher frequencies? The Nyquist–Shannon–Kotelnikov theorem implies that in order to obtain reliable information about a resonant (regular) component with frequency $f_r$, the inequality $f_d \geq 2 f_r$ must be true [Kotelnikov, 1933]. But the stochastic components are characterized by much higher frequencies than we can measure with. The purpose of this paper is to discuss how the analysis of discrete signals produced with some finite sampling frequencies may be carried out with Flicker-Noise Spectroscopy (FNS), a phenomenological framework for extracting information from stochastic series [Timashev, 2006; Timashev & Polyakov, 2007].

## 2. FNS principles

The basic idea of FNS is to treat the correlation links present in sequences of different irregularities, such as spikes, "jumps", discontinuities in derivatives of different orders, on all levels of the spatiotemporal hierarchy of the system under study as the main information carriers. It is further assumed that according to the Self-Organized Criticality (SOC) paradigm [Bak, 1997], the stochastic dynamics of real processes is characterized by intermittency, consecutive alternation of rapid changes in the values of dynamic variables on small time intervals with small variations of the values on longer time intervals. It was demonstrated that the origins of such intermittency, which occurs on every hierarchical level of the system evolution, are associated with the occurrence of complex (multiparticle, nonlinear) interactions, dissipation, and inertia [Bak, 1997].

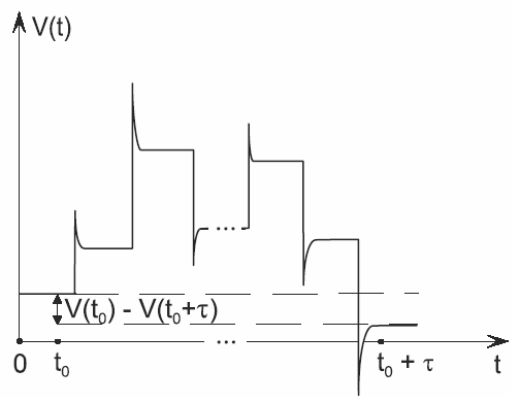

Fig. 1. Schematic of "random walk" evolution.

To illustrate the idea, consider the model process of one-dimensional "random walk" with small "kinematic viscosity" $v$ (Fig. 1). The small value of $v$ implies that when the signal changes from position $V_i$ to $V_{i+1}$, which are $|V_{i+1} - V_i|$ apart (in value) from each other, the system first overleaps ("overreacts") due to inertia and then "relaxes". We assume that the relaxation time is small compared to the residence time in a "fluctuation" position and the signal does not contain "enveloping" low-frequency curves, i.e., the resonant components are absent. The signal has all the main features of intermittent behavior: "laminar" phases with small variations in the dynamic variable $V(t)$ on characteristic time intervals $T_l$ are followed by short-term spikes in the dynamic variable on characteristic time intervals $\tau_s$ $(\tau_s \ll T_l)$. The latter ones accompany step-like changes in $V(t)$, which determine the value of the dynamic variable during the next "laminar" phase. In order to extract the information about the dynamics of the signal in Fig. 1, one may analyze the correlation links between the signal irregularities: step-like jumps in the dynamic variable, i.e., Heaviside $\theta$-functions, and inertial spikes, i.e., Dirac $\delta$-functions, accompanying the step-like changes [Schuster, 1984].

In FNS, the information parameters are assumed to be related to the autocorrelation function, one of the basic concepts in statistical physics, which is defined as

$$\psi(\tau) = \langle V(t)V(t+\tau) \rangle_{T-\tau}, \quad (1)$$

where $\tau$ is the time lag parameter, and the angular brackets denote averaging:

$$\langle (...) \rangle_T = \frac{1}{T} \int_{-T/2}^{T/2} (...) dt. \quad (2)$$

Signal $V(t)$ generally contains the "resonant" components that are specific to the evolution of the system, and the stochastic components that are related to various irregularities. The resonant modes, which may correspond to natural, external, or interferential frequencies, manifest themselves as low-frequency (slow-varying) "envelope" components of the signal. The stochastic components are characterized by wide high-frequency bands.

To extract the information contained in $\psi(\tau)$, it is convenient to analyze some transforms ("projections") of this function, for example, "power spectrum" $S(f)$:

$$S(f) = \int_{-T_M/2}^{T_M/2} \psi(\tau_1) \cos(2\pi f \tau_1) dt_1, \quad (3)$$

where $T_M \leq T/2$ is the part of the averaging interval $T$ that can be used to calculate "reliable" estimates of the actual power spectrum.

This particular transform was chosen because $S(f)$ is most effective in separating out the resonances (main components of the signal) of the analyzed functions, which are represented as a set of $N_r$ peaks characterized by positions $f_{0i}$ and "half-widths" $\gamma_i$ ($i = 1, 2, \ldots, N_r$).

The resonance contribution $S_r(f)$ to the overall power spectrum $S(f)$ can be extracted by rewriting the latter as

$$S(f) = S_c(f) + S_r(f), \quad (4)$$

where $S_c(f)$ is the continuous power-spectrum component associated with the stochastic component of dynamic variable $V(t)$. Additive representation (4) is justified because the contributions of resonant and stochastic components to dynamic variable $V(t)$ usually correspond to different time scales. In the frequency range from $1/T$ to $f_d/2$, where $f_d$ is the sampling frequency, the resonant components mostly contribute to the low-frequency range while all the irregularities manifest themselves in the high-frequency range [Timashev, 2006]. Let us note that the parameterization of $S_r(f)$ by finding the positions, "half-widths", and partial weights $A_i$ of the fixed resonances can be done rather easily. At the same time, the parameterization of $S_c(f)$ is a much more difficult task.

To solve the latter problem, we also consider difference moments ("transient structural functions") $\Phi^{(p)}(\tau)$ of different orders $p$ ($p = 2, 3, \ldots$):

$$\Phi^{(p)}(\tau) = \langle [V(t) - V(t+\tau)]^p \rangle_{T-\tau}, \quad (5)$$

where $\tau \leq T_M$.

Function (5) can also be written as a linear combination of stochastic $\Phi_c^{(p)}(\tau)$ and resonant $\Phi_r^{(p)}(\tau)$ components [Timashev, 2006]:

$$\Phi^{(2)}(\tau) = \Phi_c^{(2)}(\tau) + \Phi_r^{(2)}(\tau). \quad (6)$$

Equation (6) may be used because the contribution of resonant components to function $\Phi^{(2)}(\tau)$ is mostly seen at intermediate and large values of $\tau$. At the same time, the stochastic components contribute to the whole interval of $0 \leq \tau \leq T_M$.

The functions $\Phi_c^{(p)}(\tau)$ are formed exclusively by "jumps" of the dynamic variable while $S_c(f)$ are formed by both "spikes" and "jumps" on every level of the hierarchy, which was formally proven by Timashev [2006]. It is obvious from Fig. 1 that when the number of walks is large, the functions $\Phi^{(p)}(\tau)$ will not depend on the values of "inertial skipovers" of the system, but will be determined only by the algebraic sum of walk "jumps". At the same time, the functions $S(f)$, which characterize the "energy side" of the process, will be determined by both spikes and jumps. It should be underlined that such separation of information stored in various irregularities is attributed to the intermittent character of the evolution dynamics. In other words, the information contents of $S_c(f)$ and $\Phi_c^{(2)}(\tau)$ coincide if there is no intermittency.

## 3. Discrete time series analysis

Equations (1)-(6) were formulated for an ideal continuous series generated with an infinitely large frequency. The real series that need to be analyzed are always produced with some finite sampling frequency $f_d$. How reliable and accurate are the information parameters calculated based on the power spectrums and difference moments for the signals with that sampling frequency? Moreover, what do the mathematical notions of Dirac $\delta$- and Heaviside $\theta$-functions mean in the case of real discrete signals and how can the theory derived for continuous signals be applied to the analysis of discrete ones?

The whole frequency range for a discrete signal can usually be split into three non-overlapping ranges: (1) highest-frequency spike (high-frequency stochastic) range; (2) high-frequency jump (low-frequency stochastic) range; (3) low-frequency resonant range [Timashev & Polyakov, 2007]. In this case, the stochastic range includes both spikes and jumps, i.e., high frequencies. As $\delta$-functions may be recorded only when the sampling interval is infinitely large, in discrete signals Dirac $\delta$-functions can be interpreted as high-frequency stochastic components. It is clear that Heaviside $\theta$-functions can then be associated with low-frequency stochastic components. We can expect that when the autocorrelation function is numerically restored using the inverse cosine transform of the "power spectrum", originally calculated as the forward cosine transform of the autocorrelator, without the high-frequency spike components, any differences in the autocorrelator may only be noticed at very small values of $\tau$, which would make the contribution of the high-frequency spike components to the overall difference moment negligible. At the same time, the contribution of the high-frequency spike components to power spectrum $S(f)$ should be significant. In other words, if we calculate the power spectrum and difference moment at several different sampling frequencies, the power spectrum should change and the difference moment should keep its values almost the same. It is obvious that the spread in sampling frequencies should not be large (less than one order or so), as otherwise we would be comparing the evolutions at different scales.

Consider the example of electroencephalogram (EEG) measurements taken at the C4 electrode for a teenager with schizophrenic symptoms (marked as 545W patient) [Timashev et al., 2005]. The values of the electric potential were recorded with respect to the electrode placed on the left ear lobe of the patient. The sampling frequency was $f_d$ = 256 Hz. The signal was recorded for approximately 60 seconds and contained $N$ = 15,459 measurements (Fig. 2).

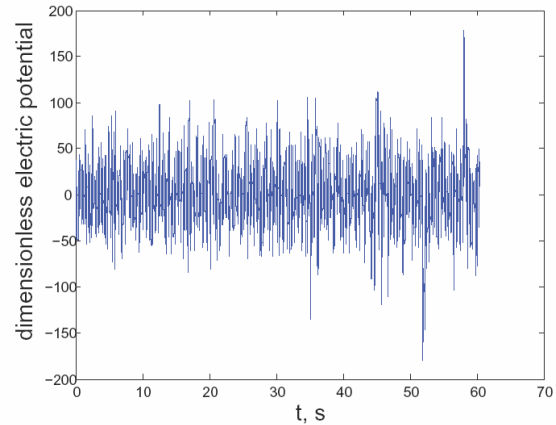

Fig. 2. Original EEG signal (C4 electrode) for 545W patient.

The power spectrum and difference moment of the second order calculated for the original signal and the signal derived by taking every fifth point from the original signal using the discrete versions of Eqs. (3) and (5) [Timashev & Polyakov, 2007] are shown in Fig. 3. In both cases, the averaging interval $T$ was set to the total duration of the measurements and the interval $T_M$ to $T/4$. The power spectrum values were normalized by dividing the power spectrum in Eq. (3) by $M = \left\lfloor \dfrac{T_M}{T} N \right\rfloor$.

It can be seen that the power spectrums are much different in the high-frequency area of Fig. 3a. The main reason is that the frequency range for the derived signal ends at 51.2 Hz and does not contain any data for higher frequencies. The high-frequency range is often associated with flicker noise, the key parameter for which is the exponent $n$ in the $1/f^n$ interpolation. A similar exponent is present in the FNS parameterization algorithm [Timashev & Polyakov, 2007]. It can be seen that the slope of the flicker-noise "tail" changes in the range of highest frequencies, which in turn changes the value of the interpolation parameter $n$. Thus, the information contents of the power spectrums displayed in Fig. 3a are different. This discrepancy is mostly attributed to the behavior of the high-frequency

"spike" components, which are the discrete "versions" of Dirac $\delta$-functions. On the other hand, Figure 3b shows only minor differences (at small $\tau$) in the values of the difference moment function calculated for both signals. It is clear that the standard deviation, an important parameter in the analysis of difference moments [Timashev & Polyakov, 2007], is practically the same for both curves. So, the difference moment appears to be stable to minor changes in the sampling interval. The same observations are valid for many other examples of real series that we have analyzed.

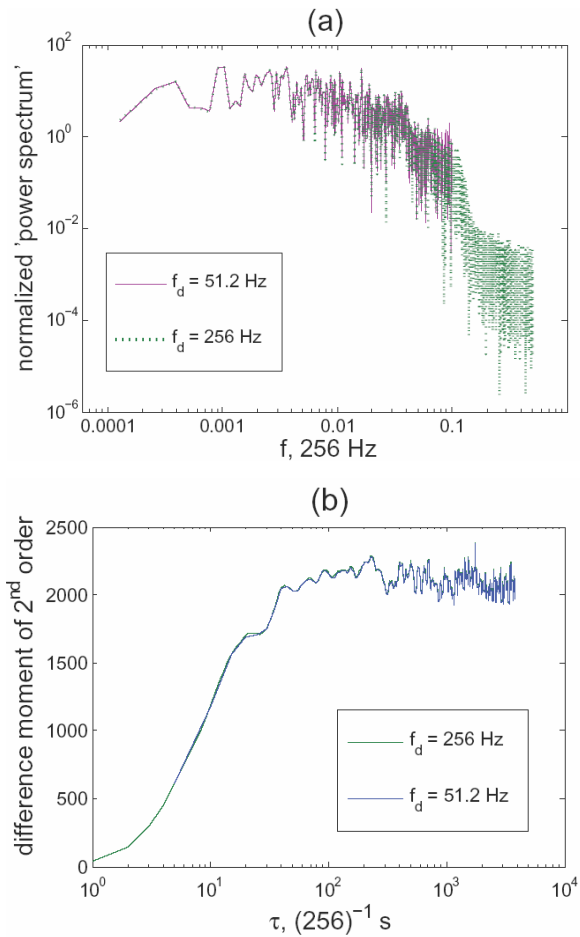

Fig 3. Power spectrum (a) and difference moment of the second order (b) for the signal in Fig. 2 and the signal derived from it by taking every fifth point.

## 4. Conclusions

Based on the above, we can conclude that:
(1) The traditional approach to extracting information from discrete signals, which is based on the Nyquist–Shannon–Kotelnikov (sampling) theorem, can be applied only to discrete signals without stochastic components. Discrete signals with stochastic components should be analyzed using phenomenological approaches like flicker-noise spectroscopy.
(2) The information contained in the power spectrum estimates and difference moments is different in the case of discrete signals, the fact that so far has been known only for continuous signals [Timashev, 2006].
(3) Changes in the sampling interval lead to changes in some information parameters. Thus the sampling interval should be considered as another parameter in general parameterization algorithms for real signals.
(4) The parameterization algorithm based on the theoretical derivations for continuous signals [Timashev & Polyakov, 2007] can be used for the analysis of real discrete signals produced with some finite sampling frequencies as it is based on the second conclusion, which is valid for both continuous and discrete signals.

This study was supported in part by the Russian Foundation for Basic Research, project no. 05-02-17079.